# Extending the mathematical palette for developmental pattern formation: Piebaldism

Michaël Dougoud[1], Christian Mazza[1], Beat Schwaller[2], Laszlo Pecze[2]

[1]Department of Mathematics, University of Fribourg, Chemin du Musée 23, CH-1700 Fribourg, Switzerland
[2]Anatomy, Department of Medicine, University of Fribourg, Route Albert-Gockel 1, CH-1700 Fribourg, Switzerland

To whom correspondence should be addressed: Laszlo Pecze, Anatomy, Department of Medicine, University of Fribourg, Route Albert-Gockel 1, CH-1700 Fribourg, Switzerland.

Tel. ++41 26 300 85 11    Fax: ++41 26 300 97 33, E-mail: laszlo.pecze@unifr.ch

# Abstract

Piebaldism usually manifests as white areas of fur, hair or skin due to the absence of pigment-producing cells in those regions. The distribution of the white and colored zones does not follow the classical Turing patterns. Here we present a modeling framework for pattern formation that enables to easily modify the relationship between three factors with different feedback mechanisms. These factors consist of two diffusing factors and a cell-autonomous immobile transcription factor. Globally the model allowed to distinguishing four different situations. Two situations result in the production of classical Turing patterns; regularly spaced spots and labyrinth patterns. Moreover, an initial slope in the activation of the transcription factor produces straight lines. The third situation does not lead to patterns, but results in different homogeneous color tones. Finally, the fourth one sheds new light on the possible mechanisms leading to the formation of piebald patterns exemplified by the random patterns on the fur of some cow strains and Dalmatian dogs. We demonstrate that these piebald patterns are of transient nature, develop from random initial conditions and rely on a system's bi-stability. The main novelty lies in our finding that the presence of a cell-autonomous factor not only expands the range of reaction diffusion parameters in which a pattern may arise, but also extends the pattern-forming abilities of the reaction-diffusion equations.

# Introduction

The various patterns on the surface of animals (fur, skin, feathers) have always been intriguing because of their diversity and presumed functions. The coloration of animal skin is due to melanin pigments that are produced by melanocytes. Melanocytes are located in the *stratum basale* layer of the skin's epidermis and in hair follicles; melanocytes secrete mature melanosomes to surrounding keratinocytes (Lin and Fisher, 2007). The localized changes in the homogeneous distribution of melanocytes or in the pigment synthesis pathway result in different patterns (Mills and Patterson, 2009). The biological function of these patterns is likely as camouflage or decoy devices rather than for communication or other physiological functions (Allen et al., 2011). Zebra stripes most probably serve as a dazzle pattern. Unlike other forms of camouflage, the intention of a dazzle pattern is not to conceal, but to make it difficult to estimate a target's range, speed, and the direction of motion (How and Zanker, 2014). Zebra stripes not only confuse big predators like lions, but also make zebra fur less attractive for flies (Egri et al., 2012).

The pattern formation mechanisms have been largely debated since the pioneering work of Alan Turing, who proposed a reaction-diffusion model to explain how very distinctive patterns may arise autonomously (Turing, 1952). Almost all natural occurring patterns can be recapitulated by this model; the seminal idea has served many different purposes (Murray, 1989; Salsa et al., 2013). However, different kinds of patterns exist that can't be recapitulated by Turing's original model, an example is the fur patterns of Dalmatian dogs. These so-called piebald patterns show randomly distributed dots of different sizes. Because of their natural irregularity, piebald patterns are not considered to be the result of cell-cell interactions based on reaction-diffusion models. The pigmented spots are the result of the stochastic migration of primordial pigment cells (melanoblasts) from the neural crest to their final locations in the skin of the early embryo (Li et al., 2011; Mort et al., 2016). However the mathematical model considering only the random movement of melanoblast do not produce sharp edges between the colored and the melanocyte-free regions (Mort et al., 2016). The spots on Dalmatian dogs appear after their birth with increasing size and contrast, indicating that other mechanisms are also involved in their formation.

The original model of Turing considers two diffusing factors, an inhibitor and an activator, which require short-range activating and long-range inhibitory effects (Turing, 1952). However, cells must translate the diffusing factor-encoded information into a biological signal. The importance of the transcription factors – as cell autonomous factors - in the pattern formation is highlighted by microbial experiments with synthetic gene circuits. The experiments revealed very versatile pattern-forming abilities of the genetically engineered organisms including the programmed bull's eye pattern (Basu et al., 2005), synchronized oscillations and travelling waves among cells (Danino et al., 2010), sequentially developing stripes (Liu et al., 2011) or rings (Cao et al.). Cell autonomous/transcription factor(s) have been considered in some mathematical studies (Marcon et al., 2016; Raspopovic et al., 2014). These models produce the classical Turing patterns with expanding range of reaction diffusion parameters.

In the following study, we present a unifying theory, how the addition of a non-diffusing transcription factor in the model allows to explaining apparently unrelated mechanisms such as Turing patterns and piebald patterns. This model is demonstrated to expand the range of pattern-forming abilities of the reaction-diffusion equations, and additionally provides a useful tool for understanding the formation of piebald patterns. We demonstrate that these piebald patterns are of transient nature However, with the scaling for the activation levels of transcription factor, we can regulate the speed of the pattern formation. If this value is close to zero, the pattern is getting frozen and it will not change significantly over time.

# Methods

## Concept

Activating (A) and inhibiting (I) diffusion factors are mobile compounds; they move rather freely in the extracellular space or between cells via gap junctions (e.g. microRNAs). The diffusion factors bind to their corresponding receptors usually localized on the surface of a cell. This induces a signaling cascade that leads to the shuttling of the TF between the cytosol and the nucleus (Cai et al., 2008; Nakayama et al., 2008; Nelson et al., 2004). If the TF is retained in the nucleus for a sufficiently long period of time, then it stimulates the translation of the "color gene", whose product acts as an attractant or activator for melanocytes (MA), and either inhibits or induces the synthesis of activator and inhibitor. This in turn results in the following four situations: I) *Activating*: the activated TF stimulates the production of both the activator (A) and the inhibitor (I); II) *Inhibiting*: the activated TF inhibits the production of both the activator(s) and the inhibitor(s); III) *Mixed I*: the activated TF promotes the production of activator(s), but blocks the production of the inhibitor(s), and IV) *Mixed II*: the activated TF inhibits the production of the activator, but activates the production of inhibitor. A schematic representation of the four situations is depicted in Fig. 1.

## Mathematical model

Let $A(t,x)$ denote the concentration of activator at time $t$ and position $x$, $I$ the inhibitor concentration, and $S$ the activity levels of TF present in the nucleus. This can also be viewed as a proxy for the expression level of the melanocyte activator/attractant (MA) within the system. We rely now on the reaction-diffusion equations for the activators and inhibitors of the following form:

$$\frac{\partial A(t,x)}{\partial t} = b_a f\big(S(t,x)\big) - d_a A(t,x) + D_a \nabla^2 A(t,x) + A(t,x) * \xi_a \quad (1)$$

$$\frac{\partial I(t,x)}{\partial t} = b_i g\big(S(t,x)\big) - d_i I(t,x) + D_i \nabla^2 I(t,x) + I(t,x) * \xi_i \quad (2)$$

where $b_a$ is the production rate, $d_a$ is the degradation rate, and $D_a$ is the diffusion coefficient of the activator, while $b_i$, $d_i$, and $D_i$ are the production rate, degradation rate, and diffusion coefficient of the inhibitor, respectively. The random variables $\xi_a$ and $\xi_a$ are distributed normally $\xi_a \sim \mathcal{N}(0, \sigma_a^2)$ and $\xi_i \sim \mathcal{N}(0, \sigma_i^2)$, in line with the study of Zheng et al. (Zheng et al., 2017). We used multiplicative white noise in the stochastic differential equation; at higher concentrations of $A$ and $I$ the absolute value of the noise is higher. This is in agreement with the experimental finding that the noise, defined as the ratio between variance and mean, increases significantly when the translation efficiency, i.e. the production rate increases (van Oudenaarden). If not noted otherwise, $\sigma_a^2 = \sigma_i^2 = 0$, i.e. there is no stochastic process in the system. The functions $f$ and $g$ enable to distinguish between the different situations of interest, i.e. I) in the *Activating* situation, $f(s) = s$ and $g(s) = s$, II) in the *Inhibiting* situation $f(s) = 1/s$ and $g(s) = 1/s$, III) in the *Mixed I* situation, $f(s) = s$ and $g(s) = 1/s$, and IV) in the *Mixed II* situation, $f(s) = 1/s$ and $g(s) = s$. The dynamics of the TFs are implemented by the following differential equation:

$$\frac{dS(t,x)}{dt} = r_v(t) \cdot \left( b_s \frac{(A/I)^2}{K+(A/I)^2} - d_s S(t,x) + r_p \right) \quad (3)$$

where $b_s$ is the rate of activated TFs transported to the nucleus, $d_s$ is the rate of TFs removal from nucleus, $r_p$ is a small rate of the activated TFs always transported into the nucleus. Thus, the value $S$ can be regarded as the degree of activation of TF, if $A \gg I$, $S \approx (b_s + r_p)/d_s$ and if $I \gg A$, $S \approx r_p/d_s$. The function $r_v(t)$ is the reaction velocity. It represents the effectiveness of the system to translate local differences in the diffusion factors into the degree of activation of TF. If not specified, we will consider it to be such that $r_v(t) \equiv 1$. Otherwise it will be assumed to be decreasing with time with maximal value 1 at time $t = 0$. In our simulations we took a Hill-type function,

$$r_v(t) = \begin{cases} 1, & t < t_0 \\ \frac{k_v^\eta}{k_v^\eta + (t-t_0)^\eta}, & t \geq t_0 \end{cases} \quad (4)$$

Where $t_0$ is the time at which the reaction velocity begins to deacrease, $\eta \geq 1$ the Hill coefficient and $k_v > 0$ the half-saturation constant.

We take the initial conditions ($t = 0$) for $A$, $I$ and $S$ to be random and uniformly distributed on the interval $J_{ini} = [0.5, 1.5]$. We also test an initial tendency in the amount of TFs (linear vertical gradient, from $S_{max}$ to $S_{min}$).

The domain of integration is a square in $R^2$ and equations (1) and (2) are treated with no flux boundary conditions. All simulations were performed in Matlab R2012b. (Mathworks Inc., MA) using finite differences. We used finite differences with a time-step set to $\Delta t = 0.01$. The reference spatial domain $[a,b] \times [a,b]$ has been discretized such that $\Delta x = \Delta y = \frac{b-a}{M}$, where the parameters usually taken for our simulations are $a = -1$, $b = 1$, and $M = 100$. The differential equations have been simulated with no-flux boundary conditions. The MATLAB code for Turing-pattern generation presented by Jean Tyson Schneider (Schneider, 2012) has been used as an initial framework for our program. All our simulations are generally stopped, when the pattern stabilizes with time (except when no steady state is reached, as will be the case when investigating piebald patterns in the Mixed I situation). More precisely, let $\varepsilon > 0$, however small and $t_0$ sufficiently large. Then we assume that the process has numerically reached steady state at time step $T_e$ defined by

$$T_e = \inf\{t > t_0 : \max_x\{|A(t-1,x) - A(t,x)|\} < \varepsilon\}. \quad (5)$$

The concept behind this definition is to let the process develop during a sufficiently large time $t_0$ and stop it, as soon as the differences between two consecutive time points are negligible.

## Model superposition

Interesting situations can also occur when multiple diffusion factors are present in the system, e.g. two activators and two inhibitors. Two mechanisms need thus to be taken into account at the same time. To this purpose let $A_1(t,x)$ be the concentration of the first activator at place $x$ and time $t$, $A_2(t,x)$ the concentration of the second activator at place $x$ and time $t$, and similarly for $I_1$, $I_2$, $S_1$, and $S_2$. All these quantities follow the following differential equations,

$$\frac{\partial A_k(t,x)}{\partial t} = b_{a,k} f_k(S_k(t,x)) - d_{a,k} A_k(t,x) + D_{a,k} \nabla^2 A_k(t,x) \quad (5)$$

$$\frac{\partial I_k(t,x)}{\partial t} = b_{i,k} g_k(S_k(t,x)) - d_{i,k} I_k(t,x) + D_{i,k} \nabla^2 I_k(t,x) \quad (6)$$

With $k = 1,2$ and where $f_k$ and $g_k$ are functions similar to the functions $f$ and $g$ in of Section b, defined according to the situations of interest. All other parameters are defined similarly as before. For the substances $S_k$, we construct them in an analogous way as before, but allow to mix the dependences to $A_k$ and $I_k$. Then the substance of interest $S$, i.e. the concentration of the melanocyte attractant, is a linear combination of $S_1$ and $S_2$, i.e. within this framework, we track

$$S = \alpha_1 S_1 + \alpha_2 S_2, \quad (7)$$

where $\alpha_1, \alpha_2$ are positive constants that we choose such that $\alpha_1 + \alpha_2 = 1$. For example to produce the patterns shown in Fig. 4A, we superpose an Activating and a Mixed I situation with

$$f_1(s) = s, \qquad g_1(s) = \frac{1}{s},$$

corresponding to the Mixed I situation, and

$$f_2(s) = s, \qquad g_2(s) = s,$$

corresponding to the Activating situation, while the substances $S_k$ are such that

$$\frac{dS_1(t,x)}{dt} = b_{s1} \frac{\left(\frac{A_1}{I_1}\right)^2}{K + \left(\frac{A_1}{I_1}\right)^2} - d_{s1} S_1(t,x) + r_p$$

$$\frac{dS_2(t,x)}{dt} = b_{s2} \frac{\left(\frac{A_2}{I_1+I_2}\right)^2}{K + \left(\frac{A_2}{I_1+I_2}\right)^2} - d_{s2} S_2(t,x) + r_p$$

# Results
## The four situations

The situation defined as "*Activating*" leads to the generation of prototypical Turing patterns. As shown in previous systems, changing one parameter produces regularly-spaced black or white dots in a background of the opposite color (Fig. 2A and 2C, respectively) and labyrinth patterns in between (Miyazawa et al., 2010). To achieve such labyrinth patterns, the inhibitor needs to diffuse faster than the activator (Fig. 2B). The involvement of concentration gradients has already been proposed before to play a role in stripe formation (Hiscock and Megason, 2015; Quininao et al., 2015). If one uses a linear gradient for the activation level of TF, the system forms regular stripes and the slope of this gradient determines the two poles (Fig. 2D). The positive pole is at the top of the slope (highest initial activity), while the negative pole is at the bottom (lowest initial activity). When starting the simulation, the first stripe appears extremely fast and close to the negative pole and the stripe generation starts from both poles. After some time, it terminates in the middle of the domain, even if the negative pole is more effective in the stripe generation.

*The Inhibiting situation* also generates regularly distributed dots as well as labyrinths; however, the diffusion coefficient of the activator needs to be higher than the one for the inhibitor, as illustrated in Fig. 2E-G. This model structure is similar to the Raspopovic model (Raspopovic et al., 2014).

The *Mixed I situation* generates the "classical" piebald patterns as observed in Dalmatian dogs (*Canis lupus familiaris*) (Fig. 3A-C). As in the activating case, modifying one parameter in the model (e.g. the production rate of the inhibitor $b_i$) results in variations of the patterns. It varies from white, randomly distributed and inhomogeneous spots (Fig. 2H) to black inhomogeneous spots on a white background (Fig. 2J). In the intermediate situation, black and white spots are relatively well dispersed (Fig. 2I). A gradient of the initial activity of the immobile TF leads to a concentration of the white spots at the bottom of the rectangular domain (negative pole), while the remaining surface (upper part) is uniformly black (Fig. 2K). Such particular patterns are commonly observed on the coat of Landseer Newfoundland dogs (*Canis lupus familiaris*) and Holstein-Friesian cattle (*Bos taurus*) (Pape, 1990). What distinguishes the *Mixed I* case from the two previous ones is the system's steady-state, when considering $D_a = D_i = 0$. Here, the mechanism leading to the pattern formation is slightly different from the original analysis of Turing; the system with $D_a = D_i = 0$ is bistable. As such, the initial conditions determine the long-term behavior of $S(t, x)$, which may attain two different states (e.g. black and white) as time increases. The random arrangement of the obtained patches results thus directly from the noisy initial conditions. The pigmented spots are the result of the stochastic migration of primordial pigment cells (melanoblasts) from the neural crest to their final locations in the skin of the early embryo (Li et al., 2011; Mort et al., 2016). Although the mathematical model considering only the random movement of melanoblast does not produce sharp edges between the colored and the melanocyte-free regions (Mort et al., 2016), the initial noisy random condition, the pre-requirement of our model, can be explained by a model incorporating random melanoblast migration with proliferation in the skin of the developing embryo (Mort et al., 2016). Adding diffusion to the modeling smoothens and destabilizes the long-term behavior of $S(t, x)$, with visible transient patterns appearing, if parameters $D_a$ and $D_i$ are sufficiently small (see S1 Movie). The spots on Dalmatian dogs appear after their birth with increasing size and contrast, indicating that a reaction-diffusion mechanisms are also involved in their formation (Fig. 3A-C). In the *Mixed II situation*, differences in diffusion rates do not play a role. Although this mechanism does not produce patterns, it generates different color tones (Fig. 2L-N).

Linear stability analyses for these four situations and the possibility to transform the three-component system into a two-component system have been performed in Supplementary material (S1 Text). Multistability in the Mixed I situation and random initial conditions lead to piebald patterns, while monostability in the Mixed II situation leads to color tones.

## The effect of initial and dynamical noise on the pattern

We distinguish initial noise (uniform distribution on a support denoted by $J_{ini}$) in the concentrations of diffusion factors and TFs, which leads essentially to different initial conditions and dynamical noise involved in the process.

We found that the initial noisy conditions had a particularly strong effect determining the final pattern, as explained in the Mixed I situation. Similarly, the initial noisy conditions almost entirely determine the precise expression of the Turing patterns in Activating and Inhibiting situations as shown in Fig. 4 upper row vs. Fig. 4 middle row, if there is only moderate noise during the process. However, strong noise has effect on the resulting pattern outcomes (see Fig4 upper row vs..fig 4 lower row). Overall, the model proposed here is thus particularly robust to small stochastic variations that may occur during chemical reactions, and relies mostly on the interacting structures between the diffusing agents and the TFs, encoded in our functions $f$ and $g$, as well as in the initial conditions

## Superposition of Turing and piebald patterns.

Performing a genetic cross of Turing- and piebald-pattern generating processes reveals a spatial overlap of the two patterns and not the mixing of patterns as shown by the fur patterns of a zebra-horse hybrid (*Equus zebra* ♀ X *Equus caballus* ♂) (Fig. 5A) or a cat (*Felis catus*) (Fig. 5B). In both animals, the pattern does not appear on the white region of the piebald patches (Fig. 5B). It is also worth noting that the background of the stripes on the hybrid animal is darker compared to the regular white-black stripes present on a zebra fur. This indicates a kind of a superposition of the two pattern-forming situations; see Fig. 5C. The presented simulation (Fig. 6D) reveals a very similar pattern formation highly reminiscent of the pattern seen on the photographed cat.

### Pattern freezing

With the parameter $r_v$, seen as a scaling for the activation levels of transcription factor, we can regulate the speed of the pattern formation. If this value is close to zero, the pattern is getting frozen and it will not change significantly over time (Fig. 6). Thus, assuming such a mechanism, a transient pattern such as the piebald pattern may last in a given form on the animal coat for a very long time. A detailed analysis of the pattern-freezing phenomenon in the *Activating situation* has been performed in another Ms. (Dougoud et al., Ms under review).

# Discussion

A third non-diffusing factor has already been considered in some previous mathematical work (Klika et al., 2012; Marcon et al., 2016; Raspopovic et al., 2014), investigating the effect of cell-autonomous factor(s) on the parameter range of reaction-diffusion equations producing Turing-type patterns. The third non-moving component is considered to be a cell membrane receptor (Marciniak-Czochra, 2003), an extracellular matrix molecule reversibly binding to diffusion factors (Korvasova et al., 2015) or a transcription factor (Raspopovic et al., 2014). Mathematical analysis of three- and also multi-component systems revealed that in the presence of a non-moving component the pattern may be formed within a wider range of the diffusion parameters; patterns may be formed with equal diffusivities of activator and inhibitor (Marcon et al., 2016), with two activators (Korvasova et al., 2015) or even with only one diffusive component (Anma et al., 2012; Marcon et al., 2016). Here we showed that the presence of a cell-autonomous factor not only expands the range of reaction-diffusion parameters in which the pattern may be formed, but also extends pattern-forming abilities of the reaction-diffusion equations.

Searching for the precise molecular identity of the model's components (*A*, *I*, *TF* and *MA*) is a challenging task for experimental work, but analyzing the existing literature allows to selecting some candidates. In the *Activating* and *Inhibiting situations,* our model produces Turing-type patterns. This type of pattern has been experimentally investigated in felids. In felids pigment-type switching controlled by *Asip* and *Edn3*, both factors acting in a paracrine way, is the major determinant of color patterns (Kaelin et al., 2012). These factors originate from the dermal papilla of hair follicles and influence the pigment synthesis pathway leading to the production of either eumelanin (brown-black) or pheomelanin (yellow-red) (Mills and Patterson, 2009). The production of *Asip* is promoted by the morphogen BMP-4 (Abdel-Malek and Swope, 2011). BMP-4 is assumed to work as an inhibitory molecule in several Turing-type developmental processes (Miura, 2007). The activatory molecule is assumed to be the bFGF diffusion factor produced by fibroblasts in the dermis. FGF and other Ras/MAPK pathway activators inhibit BMP signaling through MAPK-mediated phosphorylation of SMAD transcription factors (Sapkota et al., 2007). Besides that, the secreted protein Wnt, acting via the Wnt/β-catenin signaling pathway also inhibits BMP4 production (Baker et al., 1999). Wnt and BMP4 have been already proposed serving as a Turing pair in digit development (Raspopovic et al., 2014). Theoretically, noggin, another secreted protein, also operates together with BMP4 as a Turing-pair, because noggin i) extracellularly binds BMP4 ii) inhibits the effect of BMP4 and iii) modulates its diffusion parameters forming a noggin/BMP4 pair (McMahon et al., 1998). Molecular analysis also revealed that a transmembrane aminopeptidase, Tabulin, playing a role in the blotched tabby cat and king cheetah pattern formation (Kaelin et al., 2012). Most probably, this protein interferes with the movement of diffusion factors. Recent finding show that in African striped mice (*Rhabdomis pumilio*) and in Eastern chipmunks (*Tamias striatus*) the periodic dorsal stripes are the result of differences in melanocyte differentiation and the transcription factor ALX3 is a regulator of this process (Mallarino et al., 2016).

The localized absence of the pigment-producing melanocytes results in irregular patches of white (piebald pattern) coloration on an animal skin. Although children born with piebaldism had been displayed in circuses as "The *Zebra People*" (Huang and Glick, 2016), one needs to reconcile that the piebald patterns and zebra stripes are the result of different processes. Piebald patterns are not restricted to animals such as horses, dogs, birds, cats, pigs, and cattle, but are also observable on human skin. Piebaldism does not usually change with time. In fact, the colorless areas occasionally shrink or in some cases may spontaneously heal (Frances et al., 2015). This indicates, in line with our model, that piebald pattern is a transient pattern. One can also observe that the relative sizes of irregular patches of piebald pattern are increasing with time (see S1 Movie). If we suppose that the pattern freezing phenomenon occurs at different developmental stages of piebald patterns, this will lead to smaller or bigger patches on the animal coat. To give an example from the animal kingdom, one can compare the patterns of Appaloosa (small irregular patches) and Tobiano (big irregular patches) horses (*Equus caballus)* or the patterns of Dalmatian and Landseer Newfoundland dogs (*Canis lupus familiaris*).

The molecular background of the piebald pattern is the altered melanocyte distribution. The diffusion factor (SCF) and its receptor KIT on melanocytes were identified as factors for melanocyte survival and distribution. The clinical features and phenotypic severity of piebaldism clearly correlate with the site and the type of mutation in the *KIT* gene encoding the KIT receptor (Oiso et al., 2013; Spritz, 1994). Its ligand, the diffusion factor (SCF), produced by keratinocytes and dermal fibroblasts, also serves as a chemoattractant for melanocytes. Besides that, mutations in the diffusion factor endothelin 3 (*ET3*) and its receptor, EDNRB, have been implicated in the Waardenburg syndrome type 4, a disease characterized by piebaldism (Attie et al., 1995)**.** Interestingly, although in the normal skin the uniform melanocyte distribution is maintained and is restored after temporal destruction of melanocytes due to UV-light overexposure or moderate wound infliction (Griffiths et al., 2016), the zone with the piebald pattern does not heal as well. According to our model this may be due to the presence of a yet unidentified inhibitor present in the colorless region.

In conclusion, here we demonstrated in our modeling framework that noisy initial conditions coupled with well-defined feedback loops of the interacting factors determine, which patterns are generated. As such, piebald patterns are produced when TFs trigger the production of activators, while at the same time block the production of inhibitor. The model presented here is capable to easily reproduce current Turing patterns, while at the same time also shedding new light on phenomena such as piebaldism.

# Acknowledgements

The authors wish to thank Yannik Willing, Nina Hedvig Eriksen and Viktoria Szabolcsi for providing the photographs.

# Figures

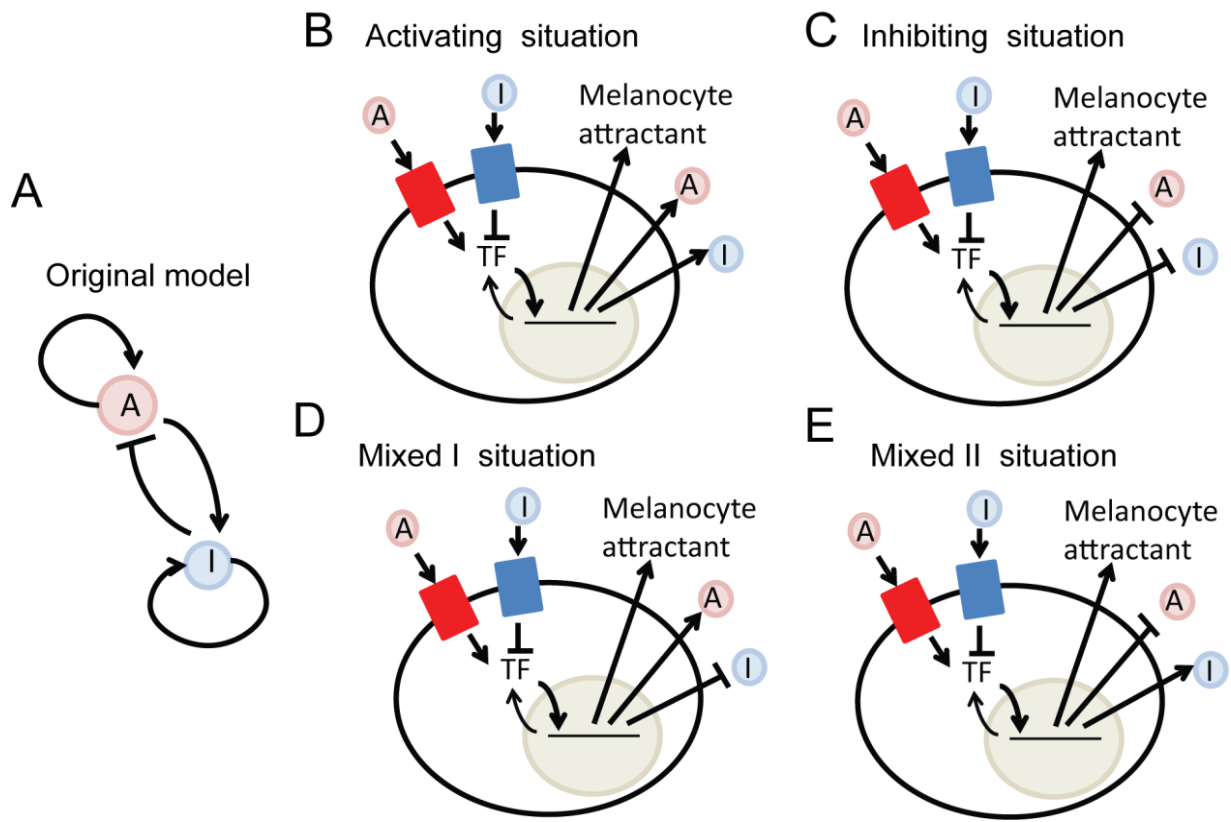

**Fig. 1. Schematic representation of the model:** **A)** The original model of Turing (Turing, 1952) considers two diffusing factors, an inhibitor (I) and an activator (A), with a requirement of long-range inhibitory effects. **B-E)** The model of the current study assumes that cells must translate the diffusing-factor encoded information into a biological signal. Diffusion factors binding to the corresponding receptors (blue and red rectangles) activating or inhibiting the transcription factor (TF). The activated TF translocates to the nucleus and I) stimulates the translation of the "color gene" acting as an attractant and II) moreover leads to the synthesis or the inhibition of the synthesis of the activator and the inhibitor. This defines four different feedback loops.

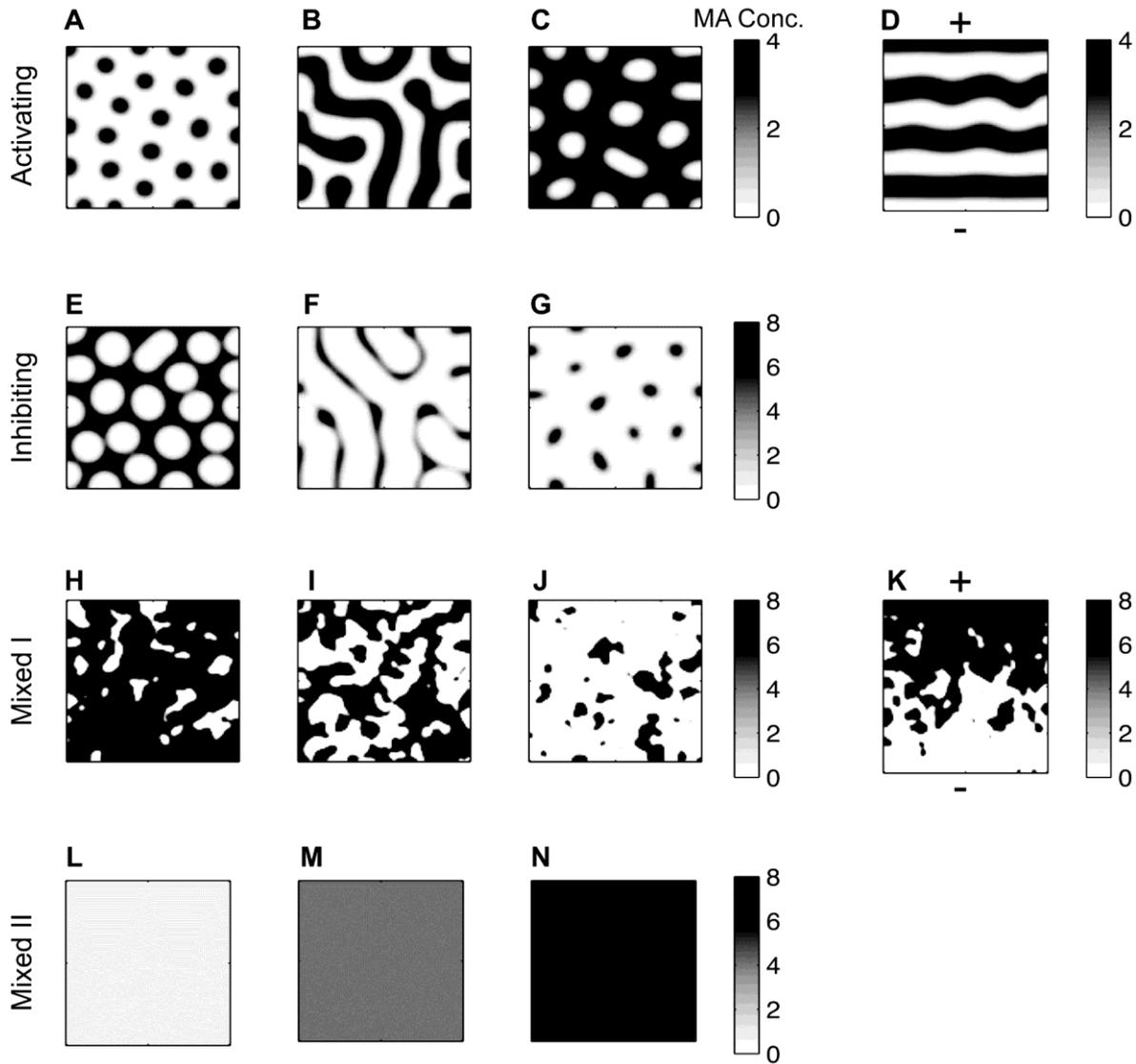

**Fig. 2. Simulations.** Initial conditions are set uniformly randomly on $J_{ini} = [0.5, 1.5]$, for $A$, $I$ and $S$ for all $x$. In all figures $S$ was plotted. *Activating situation*: "Turing patterns" i.e. regular dots and labyrinths patterns can be generated with $d_s = d_a = d_i = 0.1$, $b_s = 1$, $b_i = 0.3$, $D_a = 0.00008$, $D_i = 0.005$, $K = 50$, $\mu = 0.001$ and (**A**) $b_a = 0.5$, (**B**) $b_a = 1.1$, and (**C**) $b_a = 1.5$. In (**D**) a linear slope for the initial conditions of $S$ is used. Regular stripes begin to form at the negative pole (where there is less initially activated transcription factor). They remain regular only when close to poles. The previous parameters are used with $b_a = 1.1$ and initial conditions for $S$ set linearly along the the $y$ axis between $[0, 8]$. To obtain regular lines, the noise in $A$ and $I$ in this particular panel is reduced to $J_{ini} = [0.5, 1.5]$. *Inhibiting situation*: Regular dots and labyrinths are obtained. Parameters are similar to the previous situation (B), but $b_a = 0.9$ is fixed, $D_a = 0.005$, $D_i = 0.00008$, and $b_i$ varies such that (**E**) $b_i = 0.5$, (**F**) $b_i = 1$, and (**G**) $b_i = 1.2$. *Mixed I Situation*: Transient piebald patterns are generated. They strongly depend on the initial conditions. Here we use a large dispersal for the initial values of $A$ and $I$ ($J_{ini} = [0.5, 100]$) that enables both stable equilibria to be transiently attained with low diffusion. The parameters are similar to the previous situation, but $b_a = 1$, $D_a = 0.00006$, $D_i = 0.00001$, and $b_i$ varies such that (**H**) $b_i = 0.11$, (**I**) $b_i = 0.125$, and (**J**) $b_i = 0.15$. In (**K**) we used a linear initial gradient along the $y$-axis for $S$ with values between $[0.4, 1.6]$ and $b_i = 0.125$. *Mixed II Situation*: Grey tones are obtained. Parameters are $d_s = d_a = d_i = 0.1$, $b_s = 1$, $b_i = 0.01$, $D_a = 0.0001$, $D_i = 0.003$, and (**L**) $b_a = 0.01$, (**M**) $b_a = 1$, and (**N**) $b_a = 10$.

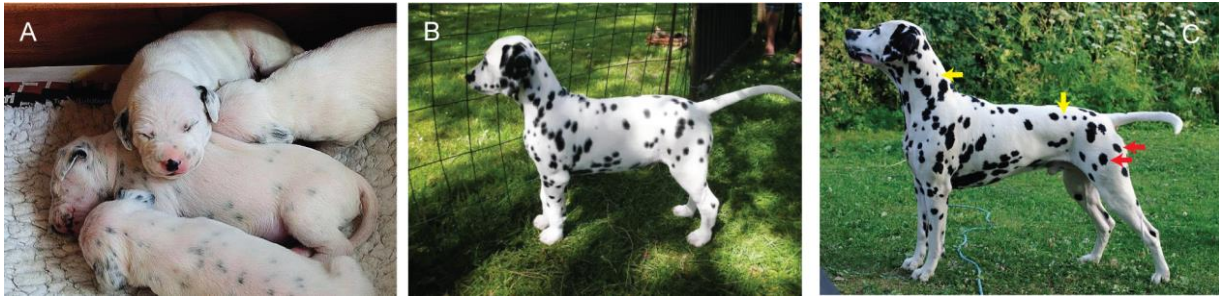

**Fig. 3**. **Developing pattern on the fur of Dalmatian dogs.** Newborn Dalmatian dogs have no or very small black dots (**A**) The same dog at age (**B**) 10 weeks and (**C**) 6 years. Newly developed dots are marked with yellow arrows. Some dots show an increase in their relative size (red arrowheads). Dalmatians were photographed by Nina Hedvig Eriksen.

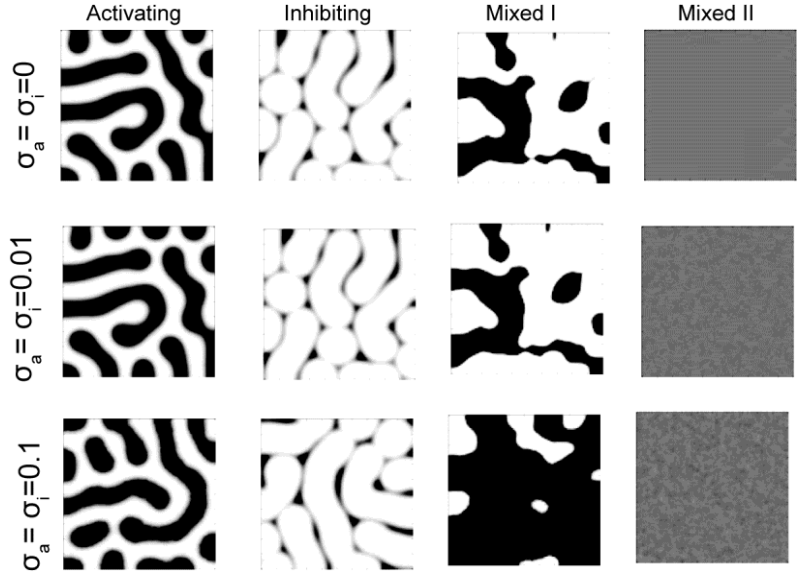

**Fig. 4**. **Robustness to noise.** The four situations are represented in each case with identical (fixed) initial conditions. In the top row $\sigma_a = \sigma_i = 0$, and there is no extrinsic noise in the system. In the middle row $\sigma_a = \sigma_i = 0.01$, thus $\xi_a$ and $\xi_b$ are uniformly distributed random variables. In the lower row $\sigma_a = \sigma_i = 0.1$, **First column**) Activating situation, with parameters as in Fig. 2B. **Second column**) Inhibiting situation, with parameters as in Fig. 2F. **Third column**) Mixed I situation, with the following parameters: $J_{ini} = [0.5, 1.5]$, for $A$, $I$ and $S$ for all $x$, $d_s = d_a = d_i = 0.1$, $b_s = 1$, $b_i = 0.375$, $b_a = 1$, $D_a = 0.0003$, $D_i = 0.000006$, $K = 50$, and $\mu = 0.002$. **Fourth column**) Mixed II situation, with parameters as in Fig. 2M.

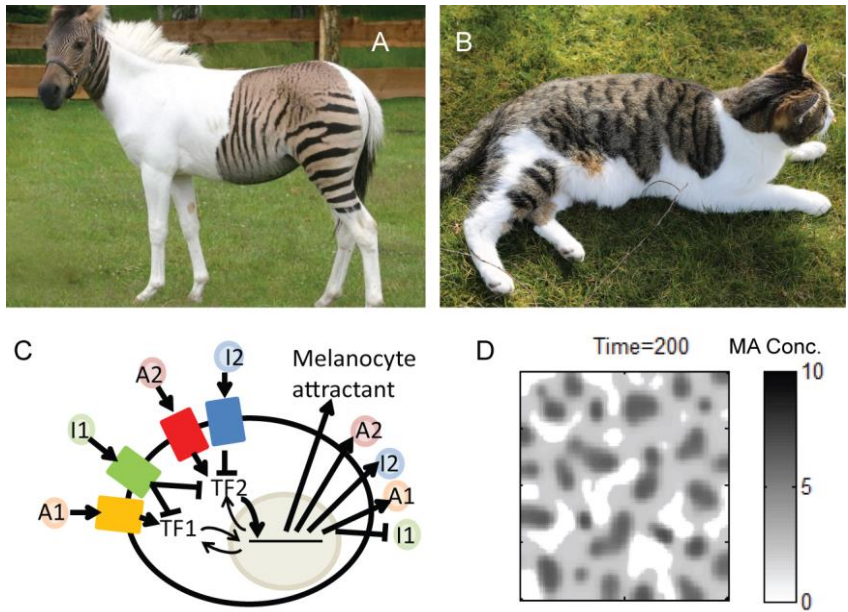

**Fig 5. Superimposed pattern:** (**A**) In a zebra-horse hybrid animal the stripes generated by a Turing mechanism appear only on the dark part of the piebald pattern (Photographed by Yannik Willing). (**B**) Cats contain both set of genes required for piebald patterns and Turing-based stripe formation. (Photographed by Viktoria Szabolcsi). (C) Schematic representation of the overlapping model. There are two sets of components; activating diffusion factors (A1, A2), inhibiting diffusion factors (I1, I2) and transcription factors (TF1, TF2) for the piebald pattern and Turing pattern generation, respectively. The two systems interact with respect to the production rate of melanocyte attractant and moreover I1 inhibits both TF1 and TF2. (**D**) Mathematical simulation. $J_{ini} = [0.5, 1.5]$, for $A_1$, $A_2$ $I_1$, $I_2$ and $S_1(0,x) \equiv 0.5$ $S_2(0,x) \equiv 0.5$ for all $x$, $d_{s1} = d_{a1} = d_{i1} = d_{s2} = d_{a2} = d_{i2} = 0.1$, $b_{s1} = b_{s2} = 1$, $b_{i1} = 0.13, b_{i1} = 0.4$, $b_{a1} = 1$, $b_{a2} = 1.1$, $D_{a1} = D_{a2} = 0.00008$, $D_{i1} = 0.000009$, $D_{i2} = 0.000009$, $K = 50$, and $\mu = 0.001$.

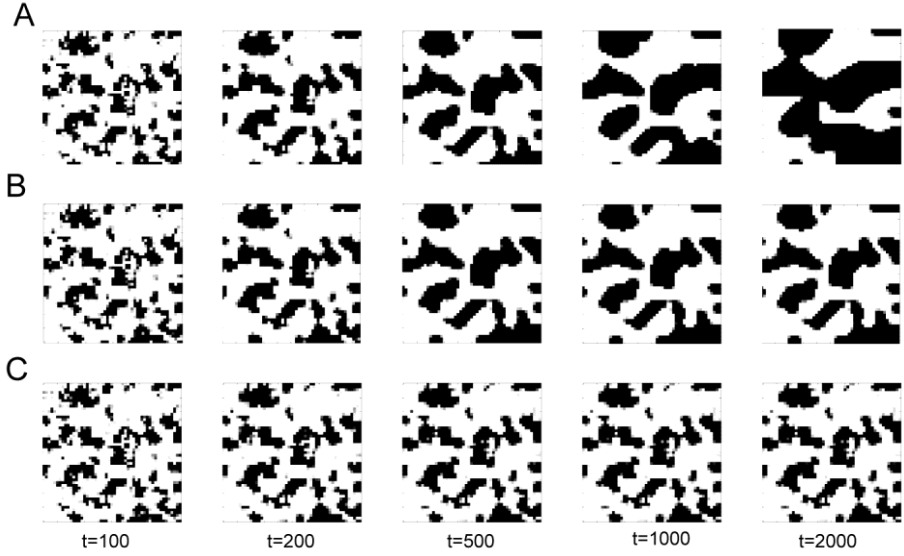

t=100   t=200   t=500   t=1000   t=2000

**Fig 6. Freezing of the piebald-pattern**. A.) The simulation is based on *Mixed I* situations with the following parameters: $J_{ini} = [0.5, 1.5]$, for $A$, $I$ and $S$ for all $x$, $d_s = d_a = d_i = 0.1$, $b_s = 1$, $b_i = 0.375$, $b_a = 1$, $D_a = 0.00015$, $D_i = 0.000003$, $K = 50$, and $\mu = 0.002$ .B) The pattern as developed in (A) is frozen with $k_v = 100$, $\eta = 3$, and $t_0 = 300$. C.) The pattern as developed in (A) is frozen with $k_v = 100$, $\eta = 3$, and $t_0 = 0$.

# Supporting information

**S1 Text. Linear stability analysis and adiabatic reduction**

**Adiabatic simplification**

Modeling explicitly the third substance $S$ has obviously its own importance on the dynamics of the process. Assuming that the corresponding chemical reactions for $S$ occur at a faster rate, one may consider a simplification of the model described in the Main Text by assuming $S$ at quasi steady state. This leads to a two-dimensional system. This model simplification consists of considering no temporal variation in $S$ and thus we have only differential equations for $A$ and $I$. As such we assume that the reactions concerning $S$ are fast with $b_s, d_s \gg 0$ and in particular larger than the other production/degradation rates of substances $A$ and $I$. From equation (3), we set

$$0 = \frac{dS(t,x)}{dt} = r_v(t) \cdot \left( b_s \frac{(A/I)^2}{K + (A/I)^2} - d_s S(t,x) + r_p \right).$$

Since $r_v(t) > 0$ for all $t$, this leads to

$$S_s(t,x) = \frac{b_s}{d_s} \cdot \frac{(A/I)^2}{K + (A/I)^2} + \frac{r_p}{d_s}.$$

This quantity is now replaced in equations (1) and (2). To this extent we need to compute $f(S_s(t,x))$ and similarly $g(S_s(t,x))$ for the different functions $f$ and $g$ considered. Denote the function of interest by $h$. For $h(s) = s$, we obtain

$$h(S_s(t,x)) = b_j \frac{b_s}{d_s} \frac{(A/I)^2}{K+(A/I)^2} + b_j \frac{r_p}{d_s}$$

with $j \in \{a, i\}$. We observe that the small random rate of TFs always transported to the nucleus $r_p$ is now transferred into the differential equation for the activator $A$ or the inhibitor $I$. Moreover, we recognize here a positive feedback of the ratio $A/I$ in the newly obtained differential equation.

Taking now $h(s) = 1/s$ leads after some simplifications to one negative and one positive feedback of the ratio $A/I$. Indeed,

$$h(S_s(t,x)) = b_j d_s K \frac{1}{r_p K + (r_p + b_s)(A/I)^2} + b_j d_s \frac{(A/I)^2}{r_p K + (r_p + b_s)(A/I)^2}. \quad (S5)$$

We illustrate this reduction in two situations. For the first, *Activating situation*, Fig. 1A represents the comparison of the pattern evolution at different time steps and with the same initial conditions. In the first row, the three-dimensional model from Section b is used, while in the second row, the corresponding model reduction is illustrated. We observe that the pattern stabilizes with time in both cases in a similar manner (Fig 1A vs.1B).

In the third, *Mixed I situation*, however, we notice a particular phenomenon that highlights very well the transient nature of Piebald patterns. In this situation, recall that $f(s) = s$ and $g(s) = 1/s$. Assuming that $S$ has reached a quasi steady state makes the process faster. As such piebald patterns tend to disappear rapidly with time as illustrated in Fig. 2B vs Fig2A. We may recover the results obtained with the two-dimensional reduction when taking the initial condition $S(0,x)$ at steady state in the three-dimensional model (Fig. 2C). To conclude and in the perspective of the application, a two-dimensional reduction works well in generating classical Turing patterns, but may fail to represent precisely piebald phenomena, because of its sensitivity to initial conditions.

To conclude and in the perspective of the application, a two-dimensional reduction works well in generating classical Turing patterns, but may fail to represent precisely piebald phenomena, because of its sensitivity to initial conditions. Again, the main effect of this reduction is the loss of the factor $r_v(t)$. With the parameter $r_v$, seen as a scaling of the activation levels of transcription factor, we can regulate the speed of the pattern formation. Moreover, such a reduction, with some parameters assumed to be large would also change the manner of estimating potentially important parameters and consequently the manner of how to view the biological phenomenon itself.

**Linear stability analysis**
We perform some further analysis to differentiate the four prototypical models by exposing their steady states and related stability properties, following a classical method (Murray, 2011). These structural properties show that the two first models (activating and inhibiting) exhibit similar behaviors, while the two mixed situations possess very different features. Recall that the systems of interest are given by equations (1-3) in Main Text which have the following form,

$$\begin{cases} \frac{\partial A(t,x)}{\partial t} = F_A(A,I,S) + D_a \nabla^2 A \\ \frac{\partial I(t,x)}{\partial t} = F_I(A,I,S) + D_i \nabla^2 I \quad \text{(S11)} \\ \frac{\partial S(t,x)}{\partial t} = F_S(A,I,S) \end{cases}$$

To perform linear stability analysis, the eigenvalues of the Jacobian $J(A,I,S)$ of the system (evaluated at steady state) will be studied in absence of diffusion. The Jacobian is the matrix of partial derivative of $F_A$, $F_I$, and $F_S$ with respect to the state variables $A$, $I$, and $S$. In other words,

$$J(A,I,S) = \begin{pmatrix} \frac{\partial F_A}{\partial A} & \frac{\partial F_A}{\partial I} & \frac{\partial F_A}{\partial S} \\ \frac{\partial F_I}{\partial A} & \frac{\partial F_I}{\partial I} & \frac{\partial F_I}{\partial S} \\ \frac{\partial F_S}{\partial A} & \frac{\partial F_S}{\partial I} & \frac{\partial F_S}{\partial S} \end{pmatrix}$$

Where every partial derivatives are evaluated at $(A,I,S)$. A fixed point of the system (S11) with no diffusion will be called linearly stable when all real parts of the eigenvalues of $J(A_s, I_s, S_s)$ are strictly negative. For simplicity, we assume here that $r_v(t) \equiv 1$ and that $d_a = d_i = d_s = d > 0$. Moreover we consider $q = 0$, i.e. $r_p = \mu$ is deterministic. We are interested in steady states. From the adiabatic simplification, we already know that

$$S_s = \frac{b_s}{d} \frac{(A_s/I_s)^2}{K + (A_s/I_s)^2} + \frac{r_p}{d}$$

independently of the situations of interest. To simplify notations, we do not write explicitly the dependence of $S_s$, $A_s$, and $I_s$ on $x$.

*I) Activating situation.* Recall that in this case we use $f(s) = g(s) = s$. As such, the system to be solved at steady state is the following

$$\begin{cases} b_a S_s - d\, A_s = 0 \\ b_i S_s - d\, I_s = 0 \end{cases}$$

Plugging $S_s$ into this system of equations leads to the following (unique) solution,

$$A_s = \frac{b_a}{d} S_s, \quad I_s = \frac{b_i}{d} S_s, \quad S_s = \frac{b_i^2 K r_p + b_a^2 \cdot (b_s + r_p)}{d \cdot (b_i^2 K + b_a^2)}$$

Now we compute the Jacobian of the system, which is given by

$$J(A_s, I_s, S_s) = \begin{pmatrix} -d & 0 & b_a \\ 0 & -d & b_i \\ \frac{2 b_s K}{(A_s^2 + I_s^2 K)^2} A_s I_s^2 & -\frac{2 b_s K}{(A_s^2 + I_s^2 K)^2} A_s^2 I_s & -d \end{pmatrix}$$

and whose eigenvalues are all equal and given by $\lambda_1 = \lambda_2 = \lambda_3 = -d < 0$ by assumption. As such the single system's steady state is linearly stable.

Now we go one step further and ask about diffusion-driven instability, as originally done by Turing in the study of his reaction-diffusion system (Turing, 1952). The basic idea is that adding diffusion in the system can produce a sufficient disturbance to generate patterns. To do so, consider the system (S11) with diffusion terms and linearized around the steady state. This problem is equivalent to the following eigenvalues/eigenfunctions problem (see Chapter 2 in Murray's book (Murray, 2011), for all details)

$$\lambda W_k = J(A_s, I_s, S_s) W_k - \begin{pmatrix} D_a & 0 & 0 \\ 0 & D_i & 0 \\ 0 & 0 & 0 \end{pmatrix} k^2 W_k,$$

where $W_k$ is the eigenfunction, $\lambda$ the eigenvalue, and $k$ is the wave number. This is thus equivalent to study the eigenvalues of the following matrix,

$$J^*(A_s, I_s, S_s) = J(A_s, I_s, S_s) - \begin{pmatrix} D_a k^2 & 0 & 0 \\ 0 & D_i k^2 & 0 \\ 0 & 0 & 0 \end{pmatrix},$$

And find some $k$, so that at least one eigenvalue has positive real part, leading to diffusion-driven instability. To do so, we will apply the Routh-Hurwitz test (See in the book entitled "An Introduction to Mathematical Biology Chapter 4.5 (Allen, 2007)). In our case, the characteristic polynomial $p(\lambda)$ is of degree three,

$$p(\lambda) = a_3 \lambda^3 + a_2 \lambda^2 + a_1 \lambda + a_0$$

Since we are interested in its roots, i.e. $p(\lambda) = 0$, we can normalize it so that $a_3 = 1$. In this case, the Routh-Hurwitz test states that the roots of $p(\lambda)$ have all negative real parts if and only if $a_2 > 0$, $a_0 > 0$, and $a_2 a_1 > a_0$. For our problem it thus suffices to find one condition that is not satisfied. After normalization, the coefficients are as follows,

$$a_2 = 3d + (D_a + D_i) k^2 > 0$$

$$a_1 = 3d^2 + 2d(D_a + D_i) k^2 + D_a D_i k^4 > 0$$

$$a_0 = d^3 + \left(d^2(D_a + D_i) - \chi(D_i - D_a)\right) k^2 + d D_a D_i k^4$$

Where we define for convenience the positive quantity

$$\chi = \frac{2 b_a^2 b_i^2 b_s d^2 K}{(A_s^2 + I_s^2 K)^2 \cdot \left(b_i^2 K r_p + b_a^2 \cdot (b_s + r_p)\right)}$$

The coefficients $a_1$ and $a_2$ are positive with our assumptions (all parameters in the system are positive). However $a_0(q)$ is a quadratic function in $q = k^2$. It is convex since $d D_a D_i > 0$. There exists thus some $k^2$ such that $a_0$ is negative whenever $a_0(q)$ possesses two roots. This is the case if the discriminant is positive, i.e. our (sufficient) condition for diffusion-driven instability is as follows,

$$\left(d^2(D_a + D_i) - \chi(D_i - D_a)\right)^2 - 4 d D_a D_i > 0$$

and the roots of $a_0(q)$ are positive (since they are of the form $k^2$). Knowing that $a_0(0) > 0$ and that $a_0$ is a parabola, we can for example impose that $a_0'(0) < 0$. Thus our sufficient condition for diffusion-driven instability is

$$\left(d^2(D_a + D_i) + \chi(D_a - D_i)\right)^2 - 4 d D_a D_i > 0$$

with

$$d^2(D_a + D_i) + \chi(D_a - D_i) < 0.$$

Note that this last condition implies $D_a < D_i$. More precisely, $D_a < D_i \frac{\chi - d^2}{\chi + d^2}$. We plotted this condition for diffusion-driven instability in Fig. S3A as function of the parameters $b_a$ and $D_a$ (all other parameters as in Fig. 2A-C from Main Text).

*II) Inhibiting situation.* Recall that in this case we use $f(s) = g(s) = 1/s$. As such, the system to be solved at steady state is as follow

$$\begin{cases} b_a \frac{1}{S_s} - d\, A_s = 0 \\ b_i \frac{1}{S_s} - d\, I_s = 0 \end{cases}$$

Plugging $S_s$ into this system of equations lead to the following (unique) solution,

$$A_s = \frac{b_a}{d} \frac{1}{S_s}, \quad I_s = \frac{b_i}{d} \frac{1}{S_s}, \quad S_s = \frac{b_i^2\, K\, r_p + b_a^2 \cdot (b_s + r_p)}{d \cdot (b_i^2\, K + b_a^2)}$$

Now we compute the Jacobian of the system,

$$J(A_s, I_s, S_s) = \begin{pmatrix} -d & 0 & -\frac{b_a}{S_s^2} \\ 0 & -d & -\frac{b_i}{S_s^2} \\ \frac{2 b_s K}{(A_s^2 + I_s^2 K)^2} A_s I_s^2 & -\frac{2 b_s K}{(A_s^2 + I_s^2 K)^2} A_s^2 I_s & -d \end{pmatrix}$$

whose eigenvalues are all equal and given by $\lambda_1 = \lambda_2 = \lambda_3 = -d < 0$ by assumption. As such the single system's steady state is linearly stable. We adopt now the same technique as for the activating situation to derive a sufficient criterion for diffusion-driven instability with help of the Routh-Hurwitz test. Let

$$\tilde{\chi} = \frac{2 b_a^2 b_i^2 b_s d^3 K \cdot (A_s^2 + I_s^2 K)}{\left(b_i^2 K r_p + b_a^2 \cdot (b_s + r_p)\right)^3} > 0$$

Then diffusion-driven instability occurs when

$$\left(d^2 (D_a + D_i) + \tilde{\chi}(D_i - D_a)\right)^2 - 4 d D_a D_i > 0$$

with

$$d^2 (D_a + D_i) + \tilde{\chi}(D_i - D_a) < 0$$

Again, this last inequality can hold only if $D_i < D_a$. More precisely, $D_i < D_a \frac{\tilde{\chi} - d^2}{\tilde{\chi} + d^2}$. Fig. S3B shows these conditions as function of the parameters $b_i$ and $D_i$ (other parameters as in Fig. 2E-G form Main Text). Note also that these sets of conditions show how similar the activating and inhibiting situations are. Their steady states are of the same nature (linearly stable without diffusion) and the conditions for Turing instability are analogous. The parameter space leading to Turing instability is larger in the inhibiting situation (compare Fig. S3 A and B).

*III) Mixed-I situation.* We will show here that the case producing piebald patterns is much richer than the two previous ones in terms of steady-states and local stability, but also trickier to handle analytically. Therefore numerical simulations are provided. In this case, the functions $f$ and $g$ are $f(s) = s$ and $g(s) = 1/s$. The system needed to be solved is

$$\begin{cases} b_a S_s - d\, A_s = 0 \\ b_i \dfrac{1}{S_s} - d\, I_s = 0 \end{cases}$$

This system possesses multiple roots, which can also be complex. The bifurcation diagram of Fig S4A shows the number of roots of this system as function of the parameter $b_i$. All other parameters are the standard ones used to produce Fig. 2H-G in Main Text; represented is the value(s) of $A_s$. For $b_i \in\ ]0\,;4.5[$ the system possesses three real roots. Interestingly, two of them (blue traces) are linearly stable, but do not have the diffusion-driven instability property. The third one is not linearly stable in absence of diffusion (even if it has the diffusion-driven instability property). To obtain such a plot, we solved numerically the system for each parameter value and tested linear stability and diffusion-driven instability with the matrices $J$ and $J^*$. In the last case we iterated the squared wave number $k^2$ until one eigenvalue was positive or some threshold attained.

Depending on the initial conditions, a particular trajectory of the process in one particular cell falls into one of the two locally linearly stable steady states while the diffusion tends to homogenize the whole pattern. This explains the transient nature of piebald patterns in our model. Fig S4B illustrates the basin of attraction of the two different linearly stable steady states. To obtain such a figure we started a one-cell process from different initial conditions $A(0)$, $I(0)$, and $S(0)$, and tracked to which steady state it converges. The parameters are the ones of Fig. 2I in Main Text and give information on the attainability of the two steady-states. The size of one basin is indeed significantly smaller than the other one. Initial conditions being (uniformly) random, the chance to reach the corresponding steady state is thus also smaller.

For larger $b_i$, the system possesses only one (stable) steady state. Such a choice would produce color tones similar to the ones observed in the Mixed-II situation.

*IV) Mixed-II situation.* In this case, the functions $f$ and $g$ are $f(s) = 1/s$ and $g(s) = s$, and the system reads

$$\begin{cases} b_a \dfrac{1}{S_s} - d\, A_s = 0 \\ b_i S_s - d\, I_s = 0 \end{cases}$$

If its form is similar to the Mixed-I situation, the solutions here consist in a single (real) steady-state, which is linearly stable. However, there is no diffusion-driven instability, as illustrated in Fig. S5 for the parameters of Fig. 2L-M of Main Text. This explains the color tones obtained within this framework.

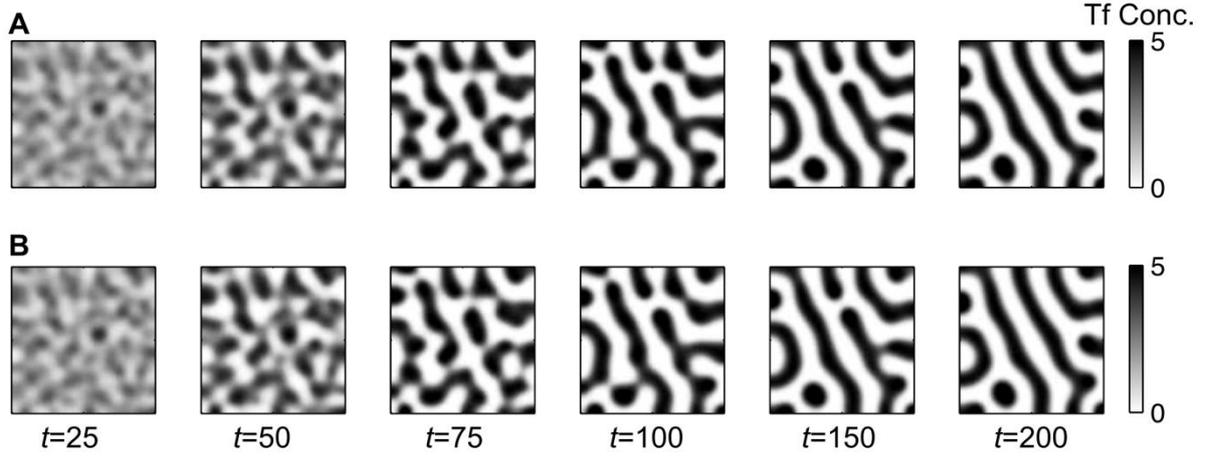

**Figure S1**. Two-dimensional and three-dimensional models show a similar behavior in the activating situation. The parameters are set as in Fig. 2B of the Main Text, except that $b_s = 20$, $d_s = 2$, and $\sigma_a = \sigma_i = 0$. (**A**) Representation of $S(t,x)$ simulated with the three-dimensional model as reported in the Main Text at different times $t$ until stabilization and with $S(0,x) \equiv 1$. (**B**) Representation of $S_s(t,x)$ simulated with the reduced two-dimensional model at the same time points and with identical (fixed) initial conditions for $A$ and $I$ as in (**A**).

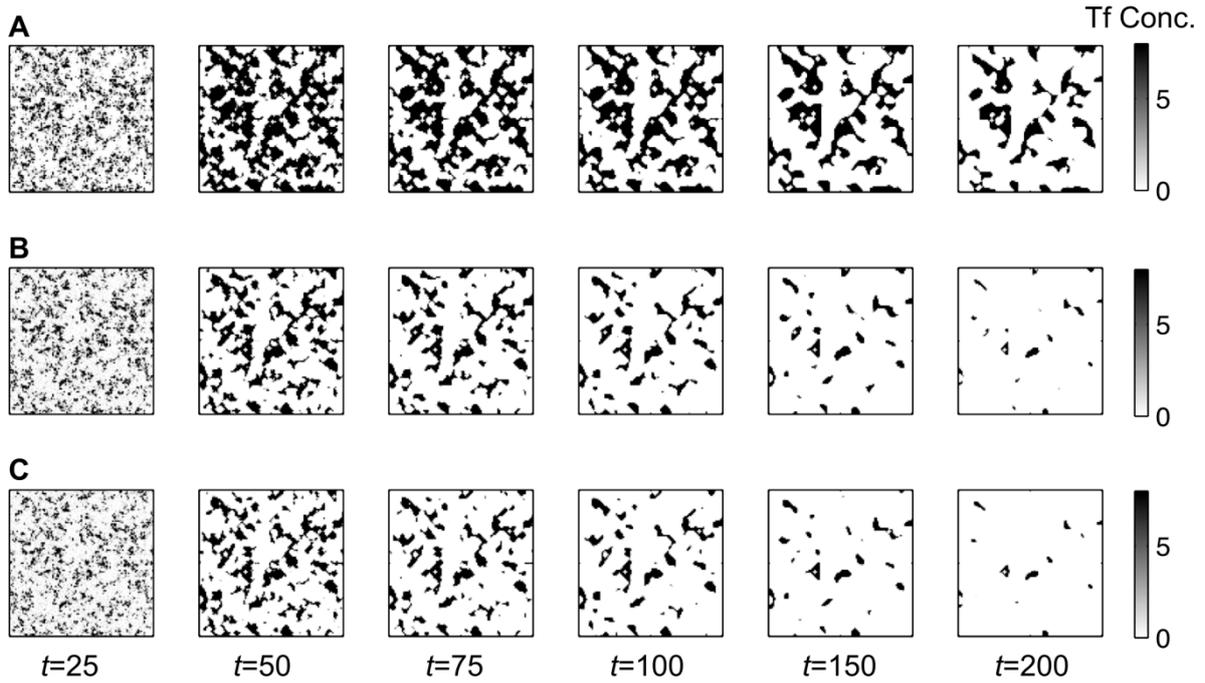

**Figure S2**. The two-dimensional and three-dimensional model show quasi-similar behaviors in the Mixed I situation with appropriate initial conditions. The parameters are set as in Fig. 2I of the Main Text, except that $b_s = 20$, $d_s = 2$, $D_a = 0.00005$, $D_i = 0.0000008$, and $\sigma_a = \sigma_i = 0$. (**A**) Representation of $S(t,x)$ simulated with the three-dimensional model as reported in the Main Text at different times points $t$ until stabilization and with $S(0,x) \equiv 1$. (**B**) Representation of $S_s(t,x)$ simulated with the reduced two-dimensional model at the same time points. The initial conditions for $A$ and $I$ are identical to the ones used in (**A**). (**C**) Same framework as in (**A**) but with initial conditions $S(0,x) = S_s(0,x)$ for all $x$ in the domain.

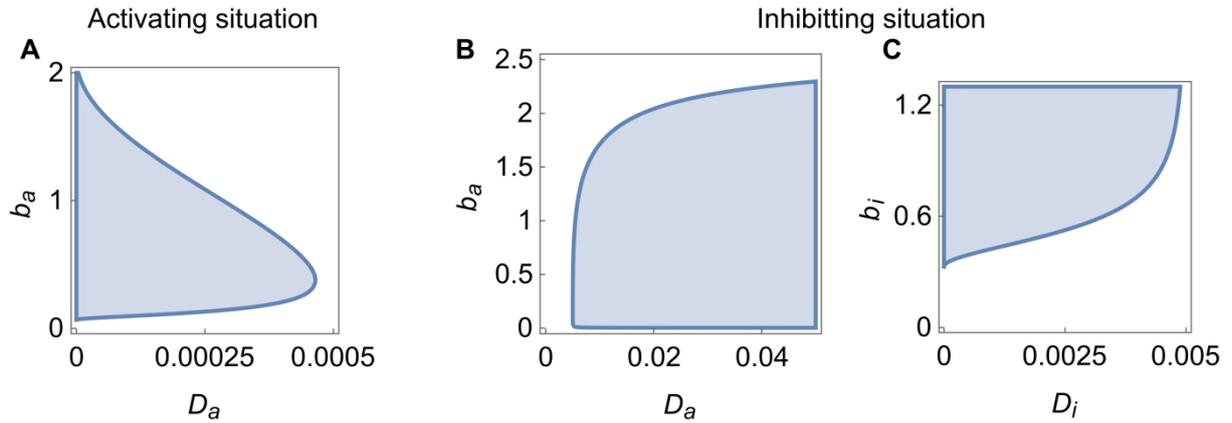

**Figure S3**. Linear stability analysis in the activating and inhibiting situations. (**A**) In the activating situation, the domain leading to the formation of Turing patterns is represented in blue as a function of the parameters $D_a$ and $b_a$; all other parameters as in Fig. 2A-C, Main Text with $\sigma_a = \sigma_i = 0$. (**B-C**) Same framework as in (A) but in an inhibiting case; see parameters of Fig. 2E-G, Main Text with $b_i = 1$ and $D_i = 0.005$ fixed in (B) and $b_a = 0.9$ and $D_a = 0.005$ fixed in (C) for a similar framework as in Fig. 2E-G from Main Text.

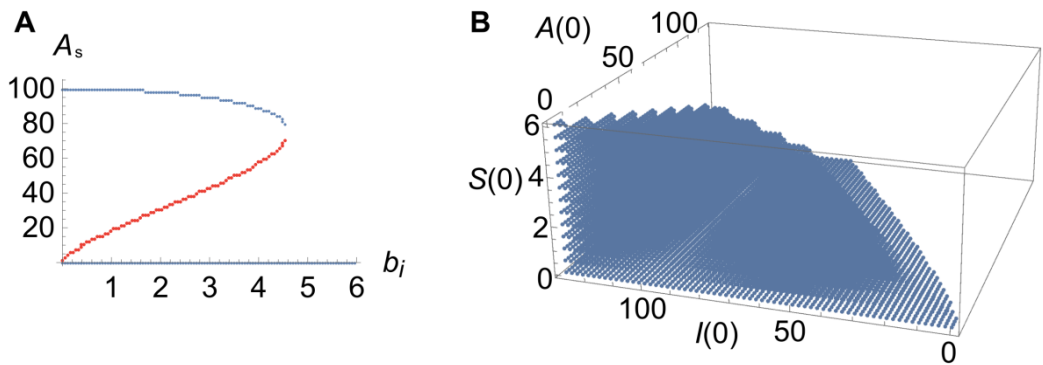

**Figure S4**. Multistability in the Mixed I situation and random initial conditions lead to piebald patterns. (**A**) Bifurcation diagram for the model leading to piebald patterns. The value of $A_s$ is computed numerically and highlighted in blue if linearly stable and red if Turing instable as a function of production rate $b_i$; all other parameters as in Fig. 2H-J from Main Text For a certain range of $b_i$ values two linearly stable steady states coexist. (**B**) Basin of attraction of the steady state $A_s = 0.1$, $I_s = 124$, $S_s = 0.01$ as a function of the initial conditions on $A$, $I$, and $S$.

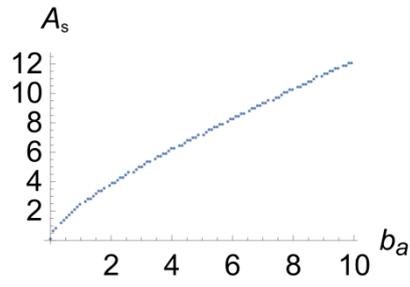

**Figure S5**. Monostability in the Mixed II situation leads to color tones. A bifurcation diagram shows here that one unique (linearly stable) steady-state is found within this framework. Parameters are as in Fig. 2L-M with $\sigma_a = \sigma_i = 0$.